# Extraordinary magnetic response of an anisotropic 2D antiferromagnet via site-dilution

Junyi Yang[1,†], Hidemaro Suwa[2,*], Derek. Meyers[3,4], Han Zhang[1], Lukas Horak[5], Zhan Zhang[6], Jenia Karapetrova[6], Jong-Woo Kim[6], Philip J. Ryan[6,7], Mark. P. M. Dean[3], Lin Hao[8,*] and Jian Liu[1,*]

[1]Department of Physics and Astronomy, University of Tennessee, Knoxville, Tennessee 37996, USA

[2]Department of Physics, University of Tokyo, Tokyo 113-8656, Japan

[3]Department of Condensed Matter Physics and Materials Science, Brookhaven National Laboratory, Upton, New York 11973, USA

[4]Department of Physics, Oklahoma State University, Stillwater, Oklahoma 74078, USA

[5]Department of Condensed Matter Physics, Charles University, Ke Karlovu 5, 12116 Prague, Czech Republic

[6]Advanced Photon Source, Argonne National Laboratory, Argonne, Illinois 60439, USA

[7]School of Physical Sciences, Dublin City University, Dublin 9, Ireland

[8]Anhui Key Laboratory of Condensed Matter Physics at Extreme Conditions, High Magnetic Field Laboratory, HFIPS, Chinese Academy of Sciences, Hefei, Anhui 230031, China

Abstract:

A prominent character of two-dimensional magnetic systems is the enhanced spin fluctuations, which however reduce the ordering temperature. Here we report that a magnetic field of only one-thousandth of the Heisenberg superexchange interaction can induce a crossover, which for practical purposes is the effective ordering transition, at temperatures about 6 times of the Néel transition in a site-diluted two-dimensional anisotropic quantum antiferromagnet. Such an unprecedentedly strong magnetic response is enabled because the system directly enters the antiferromagnetically ordered state from the isotropic disordered state skipping the intermediate anisotropic stage. The underlying mechanism is achieved on a pseudospin-half square lattice realized in the $[(SrIrO_3)_1/(SrTiO_3)_2]$ superlattice thin film that is designed to linearly couple the staggered magnetization to external magnetic fields by virtue of the rotational symmetry-preserving Dzyaloshinskii–Moriya interaction. Our model analysis shows that the skipping of the anisotropic regime despite the finite anisotropy is due to the enhanced isotropic fluctuations under moderate dilution.

Past decades have seen tremendous breakthroughs in pursuing ultrafast and securer electronics based on antiferromagnetic (AFM) materials; meanwhile, the most important challenge remains as the efficient manipulation of the AFM order parameter[1, 2]. Fundamentally, the responses of magnetic order to external stimuli are determined by spin fluctuations[3], which could be enhanced in systems featuring reduced dimensionality, spin one-half, AFM exchange coupling, and proximity to quantum phase transitions. For instance, in a two-dimensional (2D) isotropic antiferromagnet, the response to a magnetic field is primarily governed by the Zeeman energy which can be measured by a characteristic length scale $\lambda_{\text{ext}}$ given the site density. The rapid divergence of the correlation length $\xi$ as temperature decreases could lead to an extremely large field response[4, 5] when $\xi$ is comparable with $\lambda_{\text{ext}}$. On the other hand, the foremost character of spin fluctuations is their symmetry associated with magnetic anisotropy[6, 7]. The dominant impact of the magnetic anisotropy originates from the fact that, as temperature decreases, the rapidly increasing correlation length $\xi$ easily reaches the length scale $\lambda_{\text{ani}}$ characteristic of the magnetic anisotropy energy first (Fig. 1A), and the magnetic anisotropy becomes the prime perturbation to the magnetic system. As a result, easy-axis and easy-plane anisotropies necessarily lead to slower divergences of $\xi$ as temperature decreases in contrast to the strong exponential divergence in the isotropic limit ($\lambda_{\text{ani}} = \infty$) (Fig. 1A) [7]. Since the finite anisotropy in real materials usually suffice to outcompete the Zeeman field ($\lambda_{\text{ext}} > \lambda_{\text{ani}}$)[8], to create a 2D antiferromagnet with opposit length scale order field ($\lambda_{\text{ext}} < \lambda_{\text{ani}}$) will be the key to realize the large magnetic field response.

In this work, we present a proof-of-concept study showing that the AFM tunability can be systematically and significantly enhanced by exploiting magnetic dilution in the 2D antiferromagnets, giving rise to over 600% increases of the AFM onset temperature with magnetic field less than 0.5 T. Our pristine 2D antiferromagnet is realized on a pseudospin one-half square lattice embedded in a [(SrIrO$_3$)$_1$/(SrTiO$_3$)$_2$] superlattice (SL), i.e., an ultrathin film of Sr$_3$IrTi$_2$O$_9$ artificial crystal, where the neighboring IrO$_2$ planes are well separated by a nonmagnetic spacer[9]. The strong spin-orbit coupling of Ir$^{4+}$ ions stabilizes onsite $J_{eff} = 1/2$ moments[10-12] that couple antiferromagnetically via Heisenberg superexchange interactions $J$ [9, 13]. A key character of such a $J_{eff} = 1/2$ square lattice is that the large local Dzyaloshinskii–Moriya interactions $D$[14, 15], arising from the staggered IrO$_6$ octahedral rotation around the *c*-axis, tend to cancel over the lattice as a whole such that the pseudo-spins are overall almost completely isotropic – a phenomenon called hidden SU(2) symmetry (Fig. 1B)[16-19]. This further allows an in-plane uniform magnetic field $h$ to act as an effective staggered field $h \cdot \sin\varphi$, with $\tan(2\varphi) = D/J$[13, 20]. However, the high-order superexchange paths due to Hund's coupling induce a small easy-plane anisotropy and practically lower the spin symmetry from SU(2) to U(1)[18], leading to an AFM transition at ~40 K by virtue of the large $J$ and the tiny yet finite inter-plane coupling[13]. The SL is nevertheless a weakly anisotropic Heisenberg antiferromagnet in close proximity to the 2D limit.

To introduce magnetic dilution, we partially substituted Ir$^{4+}$ ions with isovalent non-magnetic Ti$^{4+}$ ions by a nominal dilution percentage $\delta$ during the atomic layer-by-layer deposition (Fig. 1B) [See supplementary for details about materials synthesis]. Figure 1C lists the XRD patterns of the SL series, where the same set of Bragg reflections can be seen, indicating that the SLs are of single-phase and epitaxially oriented along the [001] direction. The *c*-axis lattice parameter $c_{SL}$ decreases monotonically with $\delta$, consistent with the expected dependence on the volume fraction change[21], i.e., $c_{SL} = 3\delta \times c_{STO} + (1 − \delta) \times c_{pristine}$, where $c_{STO}$ and $c_{pristine}$ are the *c*-axis lattice parameters of SrTiO$_3$ and the $\delta = 0$ SL, respectively (Fig. 1D). By comparing the experimental $c_{SL}$ and the expected $c_{SL}$, we calibrated $\delta$, which turned out to be very close to the nominal values with a very small error bar [Supplementary Table. S1]. For simplicity, we use the nominal values of $\delta$ in the subsequent discussion. Note that magnetic dilution leaves the global crystal symmetry invariant [Supplementary Fig. S3].

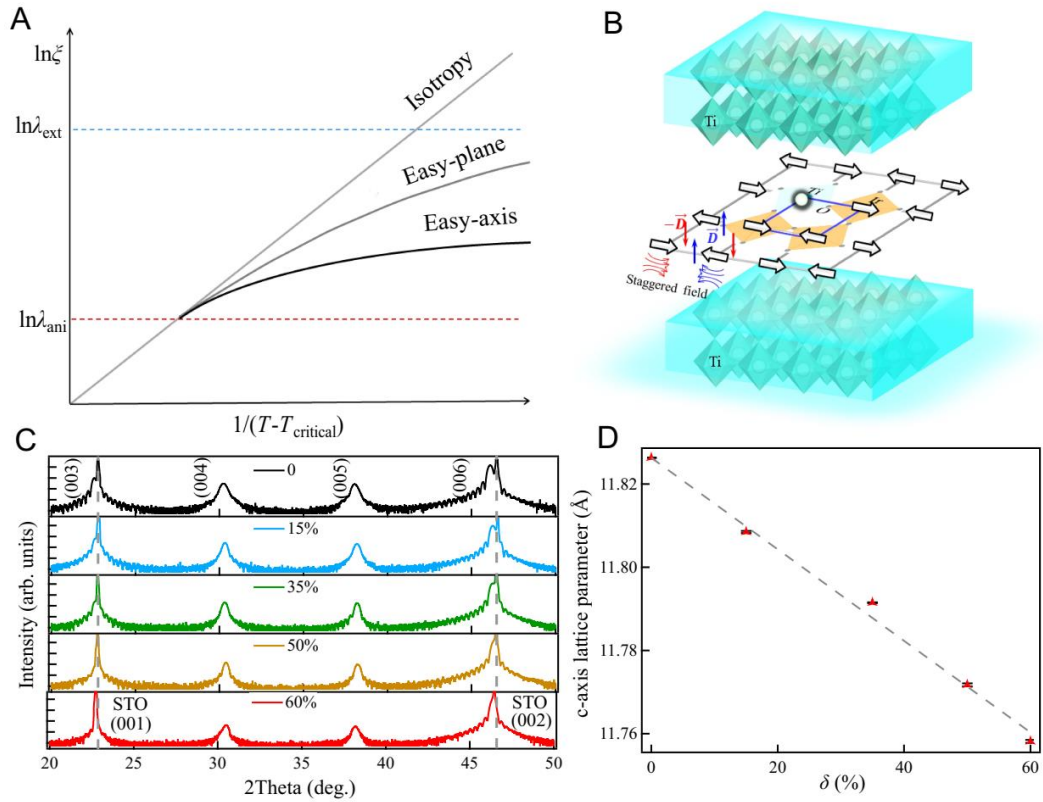

**Fig. 1. Crystal structure of diluted 2D antiferromagnets.** (**A**) Temperature-dependences of $\xi$ for different 2D magnetic systems. (**B**) Schematic diagram of $[(SrIr_{1-\delta}Ti_\delta O_3)/(SrTiO_3)_2]$ SL. (**C**) Theta-2Theta X-ray diffraction (XRD) patterns of the SLs. The SL Bragg peaks are defined in the $a \times a \times 3c$ superstructure cell, where $a$ and $c$ are the pseudo-cubic in-plane and out-of-plane lattice parameters, respectively. (**D**) $c$-axis lattice parameter, determined from synchrotron XRD around the (0 0 18) Bragg reflections [Supplementary Fig. S1], as a function of $\delta$. The error bars are deduced from experimental statistics. Dashed line denotes the expected lattice parameters. All the SLs share the same in-plane lattice parameters with the SrTiO$_3$ substrate, as confirmed from reciprocal space mapping measurements [Supplementary Fig. S2].

To further probe the AFM order in such ultrathin samples (~36 nm), we exploited magnetic resonant x-ray scattering technique. As shown in Fig. 2A, the (0.5 0.5 5) magnetic peak was observed on each SL at the base temperature, demonstrating that the AFM checkerboard type ground state[9] is preserved even under a substantial magnetic dilution. The peak intensity, which is proportional to the order parameter squared, decreases monotonically with $\delta$, which we assign to both enhanced spin fluctuations and reduced Ir$^{4+}$ content due to dilution. Figure 2C shows thermal evolutions of the magnetic order parameter (the square root of peak intensity). One can see increasing $\delta$ significantly suppresses the AFM transition, confirming the escalated AFM fluctuations. The Néel temperature $T_N$ is assigned as the onset temperature of Bragg peak and plotted against $\delta$ in Fig. 2B. Interestingly, extrapolation of the $\delta$ dependence of $T_N$ estimates that the AFM order would collapse at $\delta \sim 60\%$, which is markedly close to the theoretical percolation threshold $p_c \approx 0.593$ of a site-diluted square lattice with both nearest-neighboring and next nearest-neighboring interactions[22]. As a comparison, AFM order of a diluted square-lattice cuprate disappears around the theoretical value $p_c \approx 0.407$ with the nearest-neighboring interaction only[23]. The relatively larger next-nearest neighboring interaction between Ir sites over the cuprates has been confirmed on various iridates[24-26].

The effect of magnetic dilution is more pronounced in presence of an in-plane magnetic field. In consistency with $T_N$ defined at zero field, we define the onset temperature under nonzero magnetic field as the crossover temperature $T_0$. As shown in Fig. 2C, $T_0$ is increased by ~30% in the pristine

SL when applying a field of 0.5 T because of the effective staggered field effect in suppressing the 2D fluctuations of the AFM order[13]. When the magnetic dilution is introduced, this enhancement is doubled to ~60% at $\delta = 15\%$. It continues to increase at $\delta = 35\%$ and reaches ~600% at $\delta = 50\%$, which manifests as an extremely efficient tuning of the AFM order by suppressing the enhanced fluctuations. When comparing the energy scale, the enhanced thermal stability of the AFM order (~10 K) is at least one order of magnitude larger than the Zeeman energy of 0.5 T, highlighting the fact that the AFM order responds to the magnetic field in the nontrivial manner. Moreover, since the AFM order is stabilized by the staggered field component in the plane produced by the staggered octahedral rotation about the *z*-axis, the large magnetic response achieved under a small in-plane magnetic field also excludes the possible effect of the field enforced XY-fluctuation[27-30].

To quantify this effect, we measured the temperature dependence by systematically increasing field from 0, to 0.06, to 0.16, and finally to 0.5 T. Figures 3A-D compare the thermal evolutions on the normalized temperature scale $T/T_N$. One can clearly see that the $T_0/T_N$ is systematically increased with the applied in-plane field at all the $\delta$ values and a larger $\delta$ is beneficial for the $T_0$ enhancement under the same magnetic field. It is noteworthy that the magnetic response is especially significant at small fields and is drastically increased with increasing $\delta$. For example, 0.06 T increases $T_0$ of the $\delta = 50\%$ SL by ~12 K ($T_0/T_N \approx 3$), which is 200 times larger than the applied Zeeman energy. Figure 3E summarizes the field dependence of the normalized increase $(T_0 - T_N)/T_N$ with different $\delta$ ($T_0 = T_N$ at zero field). This quantitative comparison not only further confirms the dilution-enhanced magnetic response but also shows that the enhancement is particularly large at $\delta = 50\%$ with an extremely sharp increase of $(T_0 - T_N)/T_N$ at small fields (see Supplementary Fig. S7 for an alternative way of defining $T_N$ and $T_0$ that leads to the same conclusion). This exceptional behavior points to a possible change of the fluctuation nature under a moderate dilution.

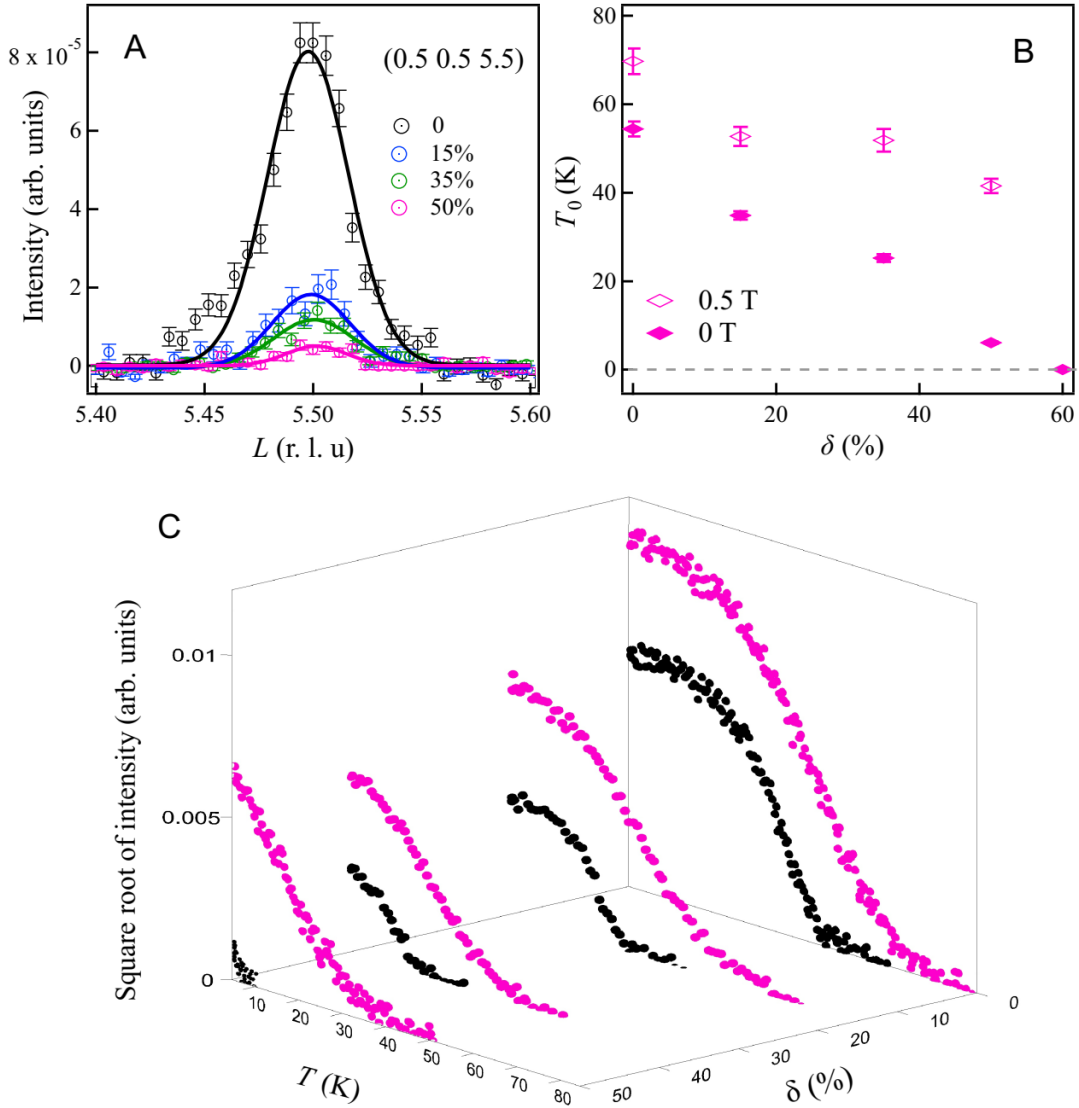

**Fig. 2. Magnetic scattering measurements on the diluted SLs.** (**A**) Base-temperature (~5 K) $L$-scans across the (0.5 0.5 5) magnetic reflection under 0 T of the SLs. (**B**) $T_N$ and $T_0$ ($B = 0.5$ T)

as a function of $\delta$. (**C**) Temperature dependent square root of the magnetic peak intensity, which characterizes the AFM order parameter, under 0 (black) and 0.5 T (pink).

To have a better understanding of the giant response in the diluted system, we conduct a theoretical anaylsis of obsered magnetic properties. To obtain an analytic expression of $T_0$, one must consider and compare the characteristic length scales of three perturbative interactions, which include the weak easy-plane anisotropy $\Gamma_1 \approx 10^{-4}$, the inter-plane coupling $|J_\perp| \approx 10^{-5}J$, and the Zeeman energy $h \cdot \sin\varphi$ [13] [Supplementary]. Firstly, in a 2D system with easy-plane anisotropy, the Berezinskii–Kosterlitz–Thouless (BKT) transition occurs at $T_{BKT}$, above which the vortex is anti-bonding and below which the vortex pair forms a bonding state [31]. The vortex is a classical solution in the continuum limit of the 2D easy-plane model. Here the vortex size is the well-defined length scale of anisotropy: $\lambda_{ani} \sim 1/\sqrt{\Gamma_1}$. Secondly, from the scaling argument [31], the crossover from the 2D system to the effective 3D system occurs when the condition $2(1-\delta)^2|J_\perp|S^2\xi^2 \sim 2\pi\rho_s$ is satisfied, where $\rho_s$ is the spin stiffness at $T = 0$, defining the length scale of inter-plane coupling $\lambda_{inter} \sim \sqrt{\pi\rho_s/(1-\delta)^2|J_\perp|S^2}$. Thirdly, the effective staggered field introduces another length scale in a way similar to the inter-plane coupling: $\lambda_{ext} \sim \sqrt{2\pi\rho_s/(1-\delta)|h\sin\varphi|S}$. Note that $\lambda_{inter}$ and $\lambda_{ext}$ can be defined regardless of spin anisotropy strength. Thanks to the 2D nature of the SL, $|J_\perp| < |h\sin\varphi|$ is achieved even at 0.06 T, leading to $\lambda_{ext} < \lambda_{inter}$ [9, 13].

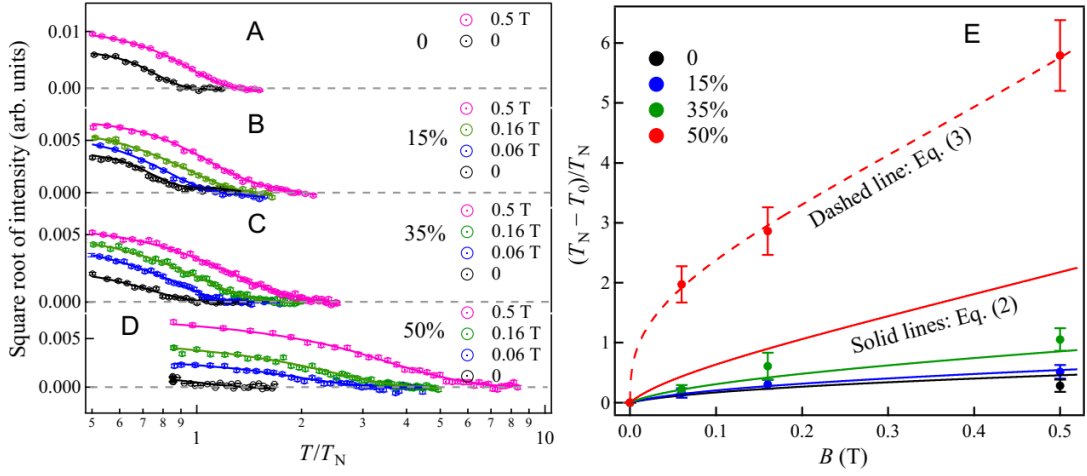

**Fig. 3. AFM response to magnetic field and theoretical analysis.** Square root of peak intensity versus $T/T_N$ for $\delta = 0$ (**A**), 15% (**B**), 35% (**C**) and 50% (**D**), under various magnetic fields. Error bars represent the statistic error. (**E**) $(T_0 - T_N)/T_N$ versus magnetic field, where $T_0(B=0) = T_N$. Solid curves are theoretical analysis for $\delta = 0$ (black), 15% (blue), 35% (green) and 50% (red) using Eq.2. For $\delta = 50\%$, simulation using Eq.3 is also shown (dashed).

In the pristine system ($\delta = 0$), we find $\lambda_{ani} < \lambda_{ext} < \lambda_{inter}$, where $\lambda_{ani} \sim 100$, $\lambda_{ext} \sim 300$ for 0.06 T, and $\lambda_{inter} \sim 400$ in units of the in-plane lattice parameter. Thus, as $\xi$ increases with decreasing temperature, the high-temperature disordered and isotropic state first crossovers to the 2D easy-plane state, where independent vortices are created (See the left panel of Fig. 4 for a schematic illustration), when $\xi \sim \lambda_{ani}$. The crossover to the 3D ordered state eventually takes place at a lower temperature such that the vortex-pair creation energy becomes comparable to the energy cost induced by $J_\perp$ and $h$ [31]. This condition when adapted for the diluted system is expressed by

$$(2(1-\delta)^2|J_\perp|S^2 + (1-\delta)|h\sin\varphi|S)\xi^2 \sim 2\pi\rho_s \ln\xi. \quad (1)$$

In the 2D system with preformed vortex, the correlation length diverges in the form $\xi \sim e^{b/\sqrt{t}}$, where $b$ is a constant depending on $\Gamma_1$ but not on $\delta$, and $t = \frac{T-T_{\text{BKT}}}{T_{\text{BKT}}}$ [31]. Using $\rho_s(T=0) \propto \rho_s(T_{\text{BKT}})$ and the Nelson-Kosterlitz relation $\rho_s(T_{\text{BKT}}) = \frac{2}{\pi} T_{\text{BKT}}$ [Supplementary], we formulate the crossover temperature of the U(1) model

$$T_0 = T_{\text{BKT}} + \frac{4b^2 T_{\text{BKT}}}{\left[\ln\left(\frac{c T_{\text{BKT}}}{2(1-\delta)^2 |J_\perp| S^2 + (1-\delta)|h \sin\varphi| S}\right)\right]^2}, \quad (2)$$

where $c$ is a constant in the prefactor of the scaling. Eq. 2 well reproduces the increase of $(T_0 - T_N)/T_N$ for all SLs with $\delta \lesssim 35\%$ as seen as the solid lines in Fig. 3(e). However, a large deviation is clearly seen for $\delta = 50\%$: the theoretical curve of Eq. 2 is significantly lower than the experimental result, indicating that the U(1) model fails in this SL. We found this conclusion very robust since our extended theoretical calculations show that this deviation is not resulted from error in $\delta$ and simply adjusting $\delta$ in Eq. 2 does not reproduce the observed response of the $\delta = 50\%$ SL [Supplementary Fig. S6]. We here argue that this is because the moderate magnetic dilution changes the picture drastically. In the diluted system, $T_N$ and $\rho_s$ significantly decrease. Accordingly, $\lambda_{\text{inter}}$ and $\lambda_{\text{ext}}$ are greatly reduced. On the other hand, as we numerically confirmed [Supplementary Fig. S5], $\lambda_{\text{ani}}$ is almost independent of $\delta$, because $\lambda_{\text{ani}}$ is well-defined in the vortex solution of the continuum limit, where the reduction of the effective coordination number is irrelevant. This is consistent with previous studies on diluted quasi-2D antiferromagnets, which reveal that the anisotropy-related critical behavior around magnetic transition is not fundamentally changed by magnetic dilution [32, 33]. Therefore, one would acquire the situation of $\lambda_{\text{ext}} < \lambda_{\text{inter}}, \lambda_{\text{ani}}$ at a sufficiently large $\delta$ even for a tiny field. The vortex-bonding picture is invalid in this regime because independent vortices are never created. Instead, the crossover emerges directly from the 2D *isotropic* state to the AFM ordered state (as shown in the right panel of Fig. 4) when the scaling relation $(1-\delta)|h\sin\varphi|S\xi^2 \sim 2\pi\rho_s$ is satisfied. $\xi \sim e^{\frac{2\pi\rho_s}{T}}$ follows the exponential divergence of the 2D Heisenberg model in the "renormalized classical" regime [34, 35]. As a result, compared to Eq. 1, the vortex-pair creation energy is replaced with the skyrmion (meron-pair) creation energy [36, 37], which is independent of $\xi$. $T_0$ in this regime is given by

$$T_0 = \frac{4\pi\rho_s}{\ln\left(\frac{a\rho_s}{d+(1-\delta)|h\sin\varphi|S}\right)}, \quad (3)$$

where $a$ and $d$ are constants that stem from the prefactor of the scaling relation and from the effect of other perturbations, respectively (Methods). The analytical forms Eq. 2 and Eq. 3 highlight the significant difference of the two regimes, i.e., the asymptotic scaling of $T_0$ in the vortex-bonding regime given by $\frac{T_0-T_N}{T_N} \sim \frac{1}{\left(\ln\frac{\rho_s}{h}\right)^2}$ is replaced by $\frac{T_0-T_N}{T_N} \sim \frac{1}{\ln\frac{\rho_s}{h}}$ in the isotropic regime, leading to a significantly larger response.

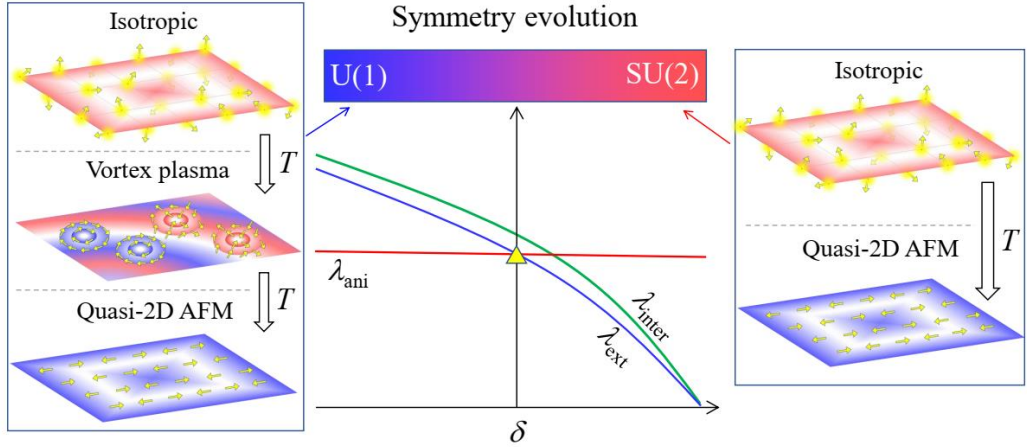

**Fig. 4. Schematic diagram of the site-dilution induced spin symmetry evolution and temperature-dependent phase change.** For $\delta$ smaller than crossing point marked by a triangle,

where $\lambda_{ani}$ is comparable to $\lambda_{ext}$, the system transitions from the high-temperature isotropic state to the quasi-2D AFM ground state via an intermediate Vortex plasma phase accounting for the U(1) spin symmetry. If $\delta$ is larger than the marked value, the isotropic state directly crossovers into the ground state unveiling an effective SU(2) spin symmetry

This regime governed by 2D isotropic fluctuations is expected to be valid for the case of $\delta = 50\%$. Using the quantum Monte Carlo method [38, 39], we obtain $\rho_s(\delta = 50\%) \approx 0.02\rho_s(\delta = 0)$ at $T = 0$, giving rise to $\lambda_{ani} \sim \lambda_{inter} \sim 100$ and $\lambda_{ext} \sim 70$ for 0.06 T. This explicitly demonstrates that we indeed achieved $\lambda_{ext} < \lambda_{inter}, \lambda_{ani}$. This switching also takes advantage of antiferromagnets, which in general have much reduced spin stiffness as compared to ferromagnetic materials. As shown in Fig. 3(e), the dashed line from Eq. 3 not only accounts for the rapid increase of $(T_0 - T_N)/T_N$ at small fields for the $\delta = 50\%$ SL but also reasonably reproduces the whole field dependence. As compared to the U(1) model of Eq. 2, the supremacy of the SU(2) model on explaining the experimental observation thus confirms the emergence of the isotropic fluctuations in the moderately diluted SL. Because $\delta = 50\%$ is approximately 10% less than the percolation threshold, this SL is expected to be outside of the quantum critical regime [23]. We emphasize that the enormous SU(2) symmetric fluctuations emerging from the highly correlated low-temperature state is realized in the 2D renormalized classical regime and thus essentially differs from the ordinary fluctuations in uncorrelated high-temperature states. Furthermore, while spin fluctuations in diluted anisotropic 2D antiferromagnets have been extensively studied [21, 23, 32, 33], the ability to switch length scales of anisotropy and external field has not been achieved, largely due to the lack of an effective staggered field effect. It can be realized in our SLs because all the necessary ingredients, including the dimensionality, the dilution, and the hidden symmetry, are implemented through top-down design and bottom-up synthesis, pointing to a new approach to studying spin fluctuations in quasi-2D magnets unobtainable in bulk synthesis.

In summary, we demonstrate a general and powerful idea of switching the order of the characteristic length scales in 2D magnets through magnetic dilution on atomic-scale. The symmetry of spin fluctuations that we can utilize by applying magnetic fields changes from U(1) to SU(2) in an anisotropic 2D pseudospin-half quantum antiferromagnet under moderate dilution. The zero-energy long-wavelength spin excitation is thus extremely sensitive to small stimuli, demonstrated here by virtue of the effective staggered magnetic field effect due to strong spin-orbit coupling. An extraordinary field-induced AFM ordering temperature increase in moderately diluted SL was observed. Our idea of the length-scale switch can be applied to easy-axis systems as well. Since dilution effects due to dopants and disorder are common in many quantum materials, these results are of intrinsic interest for both fundamental understanding and operational control of low-dimensional magnets, especially near quantum phase transitions.